\documentclass[conference]{IEEEtran}
\IEEEoverridecommandlockouts
\usepackage{cite}
\usepackage{amsmath,amssymb,amsfonts}
\usepackage{algorithmic}
\usepackage{graphicx}
\usepackage{textcomp}
\usepackage{xcolor}
\usepackage{listings}

\usepackage[disable]{todonotes} 

\def\BibTeX{{\rm B\kern-.05em{\sc i\kern-.025em b}\kern-.08em
    T\kern-.1667em\lower.7ex\hbox{E}\kern-.125emX}}

\begin{document}

\title{Triad: Trusted Timestamps in Untrusted Environments

}

\author{\IEEEauthorblockN{Gabriel Fernandez}
\IEEEauthorblockA{\textit{TU-Dresden} \\
Dresden, Germany \\
gabriel\_pereira.fernandez@tu-dresden.de}
\and
\IEEEauthorblockN{Andrey Brito}
\IEEEauthorblockA{\textit{UFCG} \\
Campina Grande, Brazil \\
andrey@computacao.ufcg.edu.br}
\and
\IEEEauthorblockN{Christof Fetzer}
\IEEEauthorblockA{\textit{TU-Dresden} \\
Dresden, Germany \\
christof.fetzer@tu-dresden.de}
}

\maketitle

\begin{abstract}
We aim to provide trusted time measurement mechanisms to applications and cloud infrastructure deployed in environments that could harbor potential adversaries, including the hardware infrastructure provider. Despite Trusted Execution Environments (TEEs) providing multiple security functionalities, timestamps from the Operating System are not covered. Nevertheless, some services require time for validating permissions or ordering events. To address that need, we introduce \textit{Triad}, a trusted timestamp dispatcher of time readings. 
The solution provides trusted timestamps enforced by mutually supportive enclave-based clock servers that create a continuous trusted timeline. We leverage enclave properties such as forced exits and CPU-based counters to mitigate attacks on the server’s timestamp counters. \textit{Triad} produces trusted, confidential, monotonically-increasing timestamps with bounded error and desirable, non-trivial properties. Our implementation relies on Intel SGX and SCONE, allowing transparent usage. We evaluate \textit{Triad}’s error and behavior in multiple dimensions.
\end{abstract}

\begin{IEEEkeywords}
Trusted Computing; Distributed Systems; TEE; Trusted Clocks;
\end{IEEEkeywords}

\section{Introduction}

Applications running within Trusted Execution Environments (TEEs) have their data privacy and integrity protected by strong properties \cite{costan2016intel, baumann2015shielding, schuster2015vc3, hunt2018ryoan, nguyen2017engarde, nguyen2016cloud}. In particular, TEEs provide a trusted environment for code execution, ensuring that the code runs strictly as intended. This guarantee makes data within the enclave challenging to tamper. Using in-enclave sources of trusted timestamps has limited usefulness since forced enclave exits prevent them from establishing a trusted, long-term continuous timeline in the presence of a malicious actor. In confidential computing applications, time intervals, i.e., relative periods to the start of the measurement, can be securely obtained, but absolute timestamps cannot, as an attacker could interrupt the application~\cite{trach2020t}. Furthermore, absolute time is also necessary for vital tasks such as ordering and comparing events. We aim to provide a cloud-based trusted timestamps service that overcomes the platform’s shortcomings and can act as a general timestamp authority for a distributed application cluster. 

A single-enclave solution cannot provide more than short-term intervals due to the impossibility of continuously executing enclave code, i.e., without operating system interruptions. Our measurements demonstrated that \textit{permanence} within the enclave is limited to small intervals before other kernel-related activities take control. Then, when the OS scheduler changes context to untrusted code, an indefinite amount of time passes, during which time-keeping threads inside the enclave do not execute. Furthermore, when the kernel halts enclave execution, an adversary may overwrite trusted timestamp counters. Some current TEEs provide the flags to verify the continuity of enclave execution, and in recent releases, an interruption handle to execute code on return \cite{sgxnotify2022}, but no means to check how long has passed.

We propose a multi-enclave, multi-node system that tackles that problem by mutually re-validating their time-keeping threads’ timestamps. Participants exchange valid timestamps to keep the system available. An external trusted timestamp source is used in highly exceptional cases to rebootstrap the system.

We tackle attacks on the in-enclave timing resources, such as the \textit{Timestamp Counter (TSC)}, by verifying its rate via CPU instruction timing and keeping in-enclave copies of the running TSC values. Our clock also provides strong properties, like strictly monotonically increasing timestamps with known error bounds.

The proposed approach leverages existing confidential computing frameworks (e.g., \textit{SCONE}~\cite{arnautov2016scone} or Graphene~\cite{tsai2017graphene}, recently renamed to Gramine), enabling us to maintain our system’s integrity. Our current implementation uses the community edition of the SCONE runtime and leverages Intel Software Guard Extensions (SGX). The SGX-powered trusted execution environment (TEE) protects the integrity and confidentiality of code and data in memory.

As contributions, we provide \textit{Triad}, an implementation of the system described above. An SGX-based system that runs on optimized Linux kernels to provide applications with trusted timestamps.

We evaluate the effectiveness of kernel configuration techniques to reduce enclave preemption rates that we use in \textit{Triad}. We also evaluate \textit{Triad} 's error performance, behavior, and corner cases.

We organize this work as follows:
In Section \ref{sec:background}, we detail related technologies and concepts. In Section \ref{sec:threatmodel}, we review the threat models for the different environments on which \textit{Triad} operates. In Section \ref{sec:approach}, we explain \textit{Triad}’s architecture and how it approaches the problem. In Section \ref{sec:properties}, we lay out the requirements we pursue and how we fulfill them. In Section \ref{sec:impl}, we describe the implementation details of the solution. We show our experiments and measurements in Section \ref{sec:eval}. In Section \ref{sec:related}, we talk about recent works in trusted timing and how \textit{Triad} differs from them. In Section \ref{sec:conclusion}, we summarize our achievements.

\section{Background}
\label{sec:background}

\subsection{Time Concepts}

This section defines relevant concepts we use to describe our solution to avoid vocabulary ambiguities.

\subsubsection{Error and Drift}
The \emph{error} between two clocks consists of the numerical difference between two timestamps or clock states in a given granularity. When referring to a single clock, \emph{error} means the difference between that clock and an established reference clock. \emph{Real-Time} refers to the time in that reference clock. \emph{Drift} is the increase in a clock’s error in relation to some other clock as time passes.

\subsubsection{Round Trip Time}
When a client reads a timestamp from a clock server via network, the request takes some time to arrive. The timestamp is sent back to the client after being read, processed, and replied to. We call the latency between the request sending and the timestamp arrival \emph{Round Trip Time (RTT)}.

\subsubsection{Clock Reading Error}

When a server reads a remote clock it assumes request and response latencies are equal and that processing time was zero. This assumption leads to the \emph{clock reading error} being set to half of the round trip time. Nevertheless, this may not be the case, specially when the network is controlled by the attacker.

\subsection{Trusted Execution Environments and Intel SGX}

Trusted execution environments (TEEs) are secure areas of computer systems (be it a memory section or a privilege level of execution) designed to protect sensitive data or binaries from tampering or unauthorized access. TEEs use hardware-based security features to create a secure environment for sensitive data in an untrusted environment. 

One such platform, Intel SGX (Software Guard Extensions), is a hardware-based security feature in Intel CPUs that provides a secure area for code and data. It creates a secure enclave, a protected area of memory isolated from the rest of the memory space. The hardware encryption engine decrypts this enclave’s memory pages when loaded into cache, outside the reach of even highly privileged users.

SGX provides users with the \textit{Timestamp Counter}. With it, threads running in an enclave can derive an interval measurement with low latency and error rates.

\subsection{SCONE}

SCONE (Secure Container Environment) is a framework for securely executing applications inside Intel SGX enclaves. It optimizes the thread model provided by SGX, reducing the costs associated with mode transitions and making it more efficient. It also simplifies the development process by providing high-level abstractions for SGX-specific functionality, relieving developers from writing low-level SGX code.


\section{Threat Model}
\label{sec:threatmodel}

This work considers three distinct domains at which different parts of the architecture rest. They are the application domain, the peer domain, and the external time service domain.

\subsection{Application Domain}

In a cloud environment, we consider a scenario where an adversary has comprehensive control over the software stack. The adversary holds sway over the cloud management system and can initiate or defer the activation or termination of virtual machines and containers, including those that house enclaves. This adversary can also shut down containers and virtual machines at their discretion.

The adversary controls all aspects of the Operating System outside enclaves. They can create, duplicate, and kill processes. Processes on non-EPC memory can be dumped and altered. Shared memory outside of the EPC and inter-process communication channels such as Linux sockets are also under their control. The adversary may cause interruptions to the execution of any application on the kernel scheduler level. 

Attacks to SGX and SCONE, such as side channel and cache-based attacks, are considered outside of the scope of this paper. We consider the adversary to be reluctant to provoke sustained periods of unavailability. While we expect that the adversary may temporarily impede progress at specific points, we expect them to allow the system to progress eventually.

\subsection{Peer Domain and External Domain}

The peer domain also sits within an untrusted environment but refers to other nodes connected by the network. That means the considerations about the Application Domain extend to the peer’s domain. Of particular importance is the ability to manipulate network packet delivery. We assume the adversary can drop, alter, delay, and repeat traffic on the network. Notice that \textit{Triad} messages are encrypted. SCONE’s Configuration and Attestation Service is responsible for key distribution and authentication. The authentication and encryption mechanisms are assumed to be computationally intractable to break. \textit{Triad} messages include nonces, which preserves freshness.

The adversary may isolate any component that relies on network communication by inhibiting all incoming and outgoing data. That applies to the external domain, where sits the \textit{external time service}, where network manipulation is possible.

\section{Approach}
\label{sec:approach}

\subsection{Overview}

We aim at providing a network-based trusted time measurement source, transparently accessed by client applications when attempting to read a timestamp from the OS. It can provide low latency and low error timestamps to trusted applications with bounded error. Client applications can trust the integrity and freshness of these timestamps.

The cluster comprises a trio of sibling servers in a local area network that uses messaging to maintain a collective notion of continuous time. When there is an enclave exit on one server, disrupting its TSC accuracy to an unknown extent, it requests one of its peers to share its current timestamp. It then proceeds to count time beginning from the received timestamp. If all nodes simultaneously get interrupted, they access an external time source the adversary cannot control. That process is undesired because these server’s latency is much higher than the latency between the peers. 

The adversary may also interfere with the rate at which the trusted clocks update. In order to detect and deter that attack, we verify TSC’s update rate against a CPU-based counting method. We calculate this method’s relationship with TSC’s initial rate during the initial \textit{calibration} protocol. 

\begin{figure}[htb]

  \centering
\includegraphics[width=\linewidth]{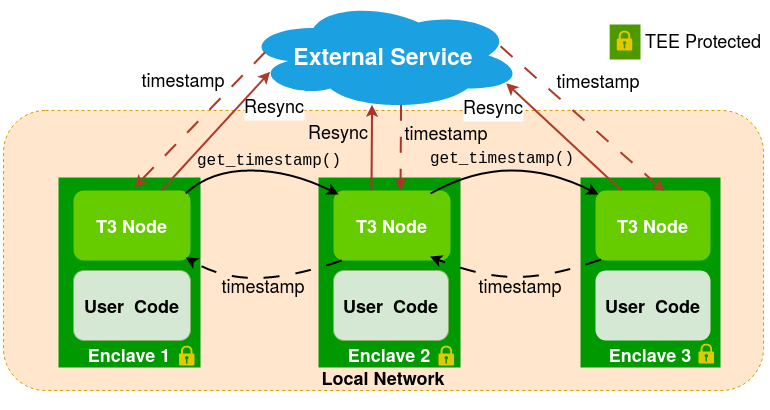}
  \caption{\textit{Triad’s} Architecture}
  \label{fig:architecture}

\end{figure}

We characterize our system as a distributed, highly available, self-correcting time-keeping mechanism with an external fail-over and synchronization time source, as portrayed in Figure \ref{fig:architecture}.

\subsection{Attack on Freshness and Drift Rate}
\label{sec:attack_freshness}

\subsubsection{Freshness}
An adversary controlling the Operating System can provoke enclave exits, thus disrupting the TSC counting. Even though the kernel’s scheduling causes them, the adversary can artificially make them long enough to increase the error of a TSC indefinitely. When control returns to the enclave thread, there is no way to assess how long (in real-time) has passed since control has been lost. An adversary can freeze a process for however long necessary to generate an exploit.

\subsubsection{Drift Rate}
\label{sec:attack_drift}
The adversary can virtualize TSC, used by the system to estimate time, making it slower. That instruction depends on an internal clock. The adversary may make access to that clock arbitrarily slow to perform this attack. That exploit is only possible when no enclaves are running, i.e., after an enclave exit.

\subsection{Preserving Freshness}

\begin{figure}[htb]

  \centering
\includegraphics[width=\linewidth]{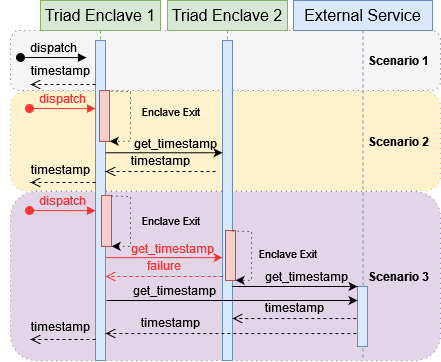}
  \caption{Failure model for Triad enclaves. One of the enclaves in the trio is omitted for simplicity}
   \label{fig:exchange}

\end{figure}

The trustworthiness of the timestamp provided is rooted in the notion of continuity of time-keeping -- which will be unpredictably but necessarily disrupted, either by the attacker or by the regular cycle of in-enclave/off-enclave execution. Each server in a trio (that composes one Triad service) continuously updates a locally cached timestamp via TSC. A server node sends its cached value to client applications that query them, as in Scenario 1 of Figure~\ref{fig:exchange}. In between updates, the servers check that they have remained inside the enclave since the last update by polling an \textit{enclave exit} flag, a.k.a., \textit{AEX}. When an exit sets the AEX, the server sets its cached timestamp to tainted.

When a timestamp cache is tainted, a node server $S_a$ blocks client requests and queries its peer servers for a timestamp in round-robin scheduling. Granted that the queried peer node $S_b$ has an untainted cached timestamp, i.e., its AEX flag was not set, it sends its cached value to $S_a$. After checking that monotonicity properties are met (as explained in subsection \ref{sec:monotonicity}), $S_a$ updates its cached timestamp with the one received from $S_b$. It then checks once again whether its timestamp is tainted and responds to blocked clients, as depicted in Scenario 2 of Figure~\ref{fig:exchange}.

When queried by a peer, a node that finds its cached timestamp also tainted responds with a failure reply and queries remaining peers for untainted timestamps. When all peers fail to reply with a timestamp, a node queries the external server for a timestamp, as illustrated in Scenario 3 of Figure~\ref{fig:exchange}.

\subsection{Confidentiality and Integrity}

Client applications access timestamps served by \textit{Triad} from inside its process space. Therefore, since they sit inside the enclave, TSC-generated timestamps (which constitute most of the total timestamps generated) cannot be read or overwritten without violating SGX assumptions.

Peer-received timestamps are enclave-encrypted. Altering bytes from the datagrams in the hopes of causing a desired change on the timestamp would be detected by the authentication mechanism. These observations apply both to the network and Operating System levels.

The candidate external time services, NTPSec \cite{raymond2016ntpsec}, and Roughtime \cite{roughtime} have mechanisms to prevent \textit{sniffing} and \textit{spoofing} attacks that compromise Confidentiality and Integrity, respectively.

\subsection{TSC Rate attack}
\label{sec:rate}

Changing the rate with which the TSC updates is a way to increase error. When the enclave is preempted, the adversary can change TSC’s rate. We use a variation of Trach’s \cite{trach2020t} CPU-instruction timing mechanism to prevent that attack. We can reliably measure periods on the millisecond magnitude by counting \texttt{ADD} instructions (see Figure \ref{fig:add_update}). By that method, we measure a small period, which we fix in 2 $ms$ to avoid error accumulation and minimize unavailability. We then compare that to the elapsed time counted via TSC. If they differ above 5\%, the runtime terminates the enclave. A new calibration protocol round must occur on the restart, adjusting the TSC rate, as explained in Subsection~\ref{sec:calibration}.
Additionally, to tackle deliberate CPU frequency changes that impact the CPU-instruction timing mechanism after enclave exits, we time main-memory access, which is not software controllable. If the latency ratio difference between main-memory access and CPU instructions goes above a more lenient $10\%$ threshold compared to what we recorded on calibration, we also terminate.  We leave the evaluation of that additional mechanism as future work.

\todo{we need to refine the architecture description to make sure there are three nodes, if there is no replication in the application, there will be empty containers that run only the time service.}

\subsection{Calibration Protocol}
\label{sec:calibration}

NTPSec and other trusted synchronization protocols do not defend against delay attacks. However, like them, our protocol must seed node servers’ TSCs and correct its value for drift. Furthermore, the adversary can manipulate TSC’s drift when the enclave is preempted.

In \textit{Triad}, we send a request to the remote server, which, upon receiving it, waits for a predefined period, which we call $PP$,  and then replies to the \textit{Triad} node. This reply must inform the client of the rate of the server compared to its own, discounting the expected round-trip time. The problem is that deliberate package delivery delay can be confounded with differences in clock rate between the remote trusted time source and the local TSC, i.e., there is no way to know which part of the latency is due to the round-trip time and which part to the local TSC being slower or faster.

Our solution consists of leveraging the time limit that control can remain in the enclave, which we call $L$. We then set $L = PP + RTT_{max}$, where $RTT_{max}$ is the maximum round-trip time we tolerate for calibration. We restart calibration if $PP + RTT_{max}$ is greater than $L$. We repeat the process for 10 seconds to improve the rate and round-trip time estimation.

The rate may be altered on every exit, as described in Section~\ref{sec:rate}. The ratio between the operation count and the TSC timing is still unknown during calibration. To address that problem, we count CPU-based instruction through the $L$ period. Calibration is complete when the number of instructions per $ms$, as counted via TSC and already corrected for drift, is established.

We shows that $L$ must sit around $1.6$ and that the calibration protocol succeeds nearly $30\%$ of the times (figure \ref{fig:enclave_length}). Certain processor operations, like TLB flushes, create a hard deadline for enclave exits and cannot be controlled by the adversary. 

Our experiments detail this data for our target architecture. We leave data portraying $L$ measurements for other architectures as future work.

\subsection{High Enclave Exit Rates and Freshness checks}

\label{sec:high_exits}

\textit{Triad} enclaves need to maintain TSC continuity throughout the sending process and for peers to verify that continuity. More precisely, the sender enclave must not have an enclave exit before sending the timestamp (as seen in Scenario 2 of Figure \ref{fig:exchange}). The peer must also not be interrupted until it receives the timestamp. To that end, it suffices that enclave execution remains uninterrupted throughout the exchange.

\subsubsection{Kernel Tunning}
\label{sec:tunning}

The longer an enclave can go on without interruptions, the fewer peer cleanups are necessary, thus amortizing clock reading error miscalculation by minimizing timestamp exchange among peers.

Similarly, unscheduled synchronization, which occurs when all server nodes have tainted timestamps simultaneously, decreases availability and may increase error, as explained in Section~\ref{sec:bounded}.

Experimental data from \cite{trach2020t} suggests that Linux systems deliver interruptions, in non-attack scenarios, at $1000\ Hz$, limiting the mean in-enclave interval to $5\ ms$. 

We mitigate the impact of system interrupts by applying several modifications to kernel configuration. These included lowering the timer interrupt frequency to its minimum ($100\ Hz$), enabling the dynamic ticks kernel mechanism, and directing all device interrupts to core 0. We took this opportunity to attempt to verify \cite{trach2020t}'s conclusions, as seen further in \ref{sec:enclave_exits} at Figure \ref{fig:aex_comb}.

\todo{Evaluate the impact of the timer for when we will suggest a new type of "cloud node" and instance family. This is a good contribution for a thesis.}

\subsubsection{TSC Register Overwrite Resistance}

A strawman approach to serving timestamps would be to calculate them based on a drift correction and an on-demand read of the TSC, i.e., once a client or peer makes a request, the server would read the TSC timestamp and apply the drift rate found during calibration to it. However, the registers that contain a TSC value are vulnerable to being overwritten by an adversary on enclave exit. Before serving a cached timestamp to a client, a \textit{Triad} server node would check if that cached timestamp is tainted and, if so, perform an untainting operation by requesting a timestamp from a peer.

In that solution, the last cached timestamp would be the last served to a client. If client requests were sufficiently far in between, the last cached timestamp would always be lower than the incoming one from the peer. As explained further in Section~\ref{sec:monotonicity}, a server node ignores received timestamps lower than its last cached timestamp. In non-attack scenarios, timestamp untainting from peers is a source of error due to clock reading error miscalculation, originating from network unpredictability (as seen in Figure~\ref{fig:error}). Therefore, the most desirable is the \textit{self-untainting} operation, i.e., receiving a lower timestamp from a peer, ignoring it and using the last cached timestamp. To maximize the rate with which a node’s timestamp is higher than one received from a peer, we keep a dedicated thread to update the cached timestamp with a TSC read multiple times per millisecond.

\section{Properties}
\label{sec:properties}

\subsection{Trusted Time Requirements}

We propose requirements (R1) and (R2) to establish a reliable and consistent trusted time clock within the distributed system. The first requirement (R1) ensures that the maximum error between the clock’s trusted time value and the real-time is bounded by the parameter $\epsilon$. That means that events associated with a timestamp cannot have happened arbitrarily later or earlier than the moment the timestamp describes. In more formal terms:

\centerline {$\forall t \in Time, \forall T \in Threads:  | TT_T(t) - t | \le \epsilon$ \ \ \ (R1)}

\todo{How much is epsilon? We never evaluate the error under pressure/attack?}

The second requirement (R2) guarantees that the trusted time values are strictly increasing, allowing for correct event ordering and ID generation.

To support (R2), it is essential to ensure that the trusted time provided by the clock is strictly monotonic. In other words, as time progresses, the trusted time values returned by the clock should always increase. We can define this property as follows:

$
\forall t_1, t_2 \in Time, \forall T \in Threads:  \\ t_1 < t_2 \Rightarrow TT_T(t_1) < TT_T(t_2) \ \ \ (R2)
$

Requirement (R2) guarantees that the trusted time values will always maintain a consistent ordering. It ensures that events occurring later will have a higher trusted time value than events occurring earlier.

Distributed systems can maintain synchronization, enforce causality, and provide reliable temporal information for various operations and processes with a trusted time clock that adheres to these properties.

\subsection{Monotonicity}
\label{sec:monotonicity}

Maintaining property (R2) is not possible when a server node $T_T$ reads a timestamp older than its last cached timestamp, i.e., $TT_T(t) < TT_U(t + \Delta)$, with $\Delta$ being the time duing which control was out of the enclave.

Intuition suggests that the server node should reject timestamps that do not comply with property R2 and query another peer. However, there is necessarily one server node that has the highest timestamp:

\centerline{$\forall t \in Time, \forall T, U \in Threads: \exists TT_T \ni TT_T(t) > TT_U(t)$}

This server would not make its cached timestamp untainted from a peer’s timestamp and would have to resort to undesired access to the external time service.

Furthermore, this solution poses two additional problems: positive error tends to accumulate, as servers with higher cached timestamps will be preferred, and error due to clock reading error may increase as peer reads happen more often.

Therefore, if a timestamp received from a peer is lower than the last cached timestamp after an exit, we increase the previous timestamp by the lowest unit our resolution permits. We then untaint our last cached timestamp, i.e., unset the AEX flag. Incrementing the timestamp maintains (R2), thus the timestamp’s ordering property. It also decreases clock reading error miscalculation, prevents positive error creep, and decreases access rates to the external time service.

\subsection{Bounded Error Timestamps}
\label{sec:bounded}

While network messaging can be blocked and the adversary can preempt enclave execution, \textit{Triad} limits timestamp’s error $|TT_T(t) - t|$ to a maximum $\epsilon$. An event that occurred during the time during which control was outside of the enclave has to have happened between the last cached timestamp $TT_T(t - \Delta)$ (which is recorded immediately before preemption, as explained in Section~\ref{sec:monotonicity}) and the timestamp received from a peer or external time service after enclave execution has returned. We describe maximum error $\epsilon$ for a timestamp requested during off-enclave time as:

\centerline{$\epsilon = \Delta + RTT$}

With:

\centerline{$RTT = RTT_1 + RTT_2$}

where $\Delta$ is the time control outside the enclave and $RTT_1$ and $RTT_2$ are the first and second message latencies of the round trip time, i.e., $|RTT_1 - RTT_2|$ describes the clock reading error. In this case, the maximum $RTT$ is such that $min(RTT_1, RTT_2) \approx 0$, and therefore $max(RTT_1, RTT_2) \approx RTT$.

The potential clock reading error, which is at most $RTT$ and at least $RTT/2$, is the main contributor to an increase in $\epsilon$ in non-attack scenarios -- where $\Delta \approx 0$. Thus the need to avoid the external time service update.

\subsection{Time Guarantees Model}

In his work, Alder et al. \cite{alder2023time} propose the characterization of trusted clocks, specifically in the field of Trusted Execution Environment-based clocks, within categories of increasingly robust properties. $T_1$ timers produce only monotonically increasing timers. The $T_2$ tier represents remote but completely trusted clocks. This tier’s caveat is that the adversary can intercept, delay, and replay timestamps.
In tier $T_3$ timers, the adversary must actively cause an interrupt to delay them. The authors call this tier \textit{Local Time}.
Tier $T_4$ refers to timers whose clocks are trusted and are generated synchronously -- the cannot be interrupted or delayed.

We argue that, despite \textit{Triad} timestamps coming, often, from remote sources, such as the external time service (which exemplifies $T_2$ well), we can characterize it as a $T_3$ timer.
While \textit{Triad} locally generated timestamps are intuitively inscribed in this tier, remote ones are more nuanced. An unbounded wall-clock time may pass while the adversary delays a timestamp message. This unbounded time, however, does not imply that timestamps have arbitrary error since we can identify its bounds, i.e., $\epsilon$ is known. By checking the state of the AEX flag after remote-read operations, we effectively turn them into transactions, which prevents timestamps from being delivered outside of continuous enclave execution. The consequence of impeding the conclusion of the transactions by generating enclave exits is that error related to the elapsed time of the operation increases, but error in relation to the commit time does not. Applications must be aware of that difference.

\section{Implementation}
\label{sec:impl}

\subsection{\textit{Triad} Runtime TEE Framework and Communication}

\textit{Triad} is implemented in Rust version 1.64.0, which was the latest compatible with SCONE version 5.8. We have chosen SCONE as our TEE framework due to the facility of running Rust code inside SGX enclaves and the ability to intercept system calls while maintaining a minimal EPC footprint.

TLS and other TCP-based protocols perform re-sends and ordering verification steps that are undesirable for applications where lowering latencies is imperative. UDP has shown to be a better alternative. We guarantee message privacy and integrity by encrypting datagrams with symmetrical AES-GCM-256 encryption.

\subsection{Architecture}

In this work, \textit{Triad} is chiefly used as an application-specific library. This mode is a transparently activated when Linux’s \texttt{get\_time} primitive is called. We achieve this by modifying SCONE’s runtime to intercept these calls and querying the \textit{Triad} service instead of forwarding it to the kernel. On initialization, it is necessary that \textit{Triad} has network access to the other participant server nodes, as per the configuration. Therefore, the other application instances must be running on the custom kernel machines for the service to be available. Because both the \textit{Triad} service and the application thread reside on the same physical node and enclave, the client application has memory space access to \textit{Triad} timestamps. In this mode, attacks on communication channels between the client and its local \textit{Triad} node are impossible within SGX’s threat model.

On the cloud level, \textit{Triad} runs on a specific instance \emph{flavor} that provides the low-enclave-exit custom kernel. It also runs within a cluster configuration, e.g., a Kubernetes’s Deployment or DaemonSet, that instantiates the complete set of nodes. We use a trio of nodes as our default architecture. This choice guarantees at least one degree of fail-over because peer clean ups are impossible with a single node. The trio formation also decreases the chances of external time service updates. Applications without replication still need to instantiate two additional nodes (using \textit{Triad} standalone images) to setup the minimum cluster.

\todo[inline]{(future work, here or thesis) Define exactly the architecture for time enabled instances}

\section{Evaluation}
\label{sec:eval}

\subsection{CPU-based Timing and In-Enclave Period Limit}
\begin{figure}[htb]
  \centering
  \includegraphics[width=\linewidth, height=5cm]{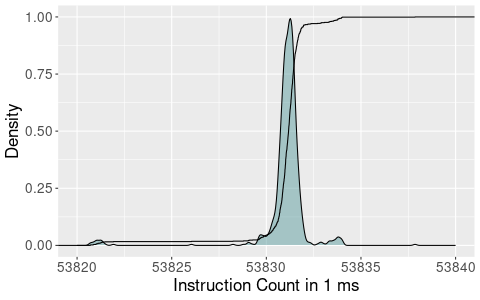}
  \caption{Instruction count in 1 millisecond. All trials concentrate on a small range.} 
  
  \label{fig:add_update}
\end{figure}

This experiment tests our CPU-based timing precision, used to monitor changes in the TSC rates on return to enclave execution. We count the number of \texttt{ADD} instructions per $ms$ as measured by an NTP-based reference clock in $1 \times 10^6$ trials. Figure~\ref{fig:add_update} shows the distribution of operation counts. 

Note that the accumulated distribution line goes from $0.00$ to $1.00$ within the range of 20Hz, from a total mean count of $53830 Hz$, as indicated by the accumulated distribution line.

\begin{figure}[htb]
  \centering
  \includegraphics[width=\linewidth]{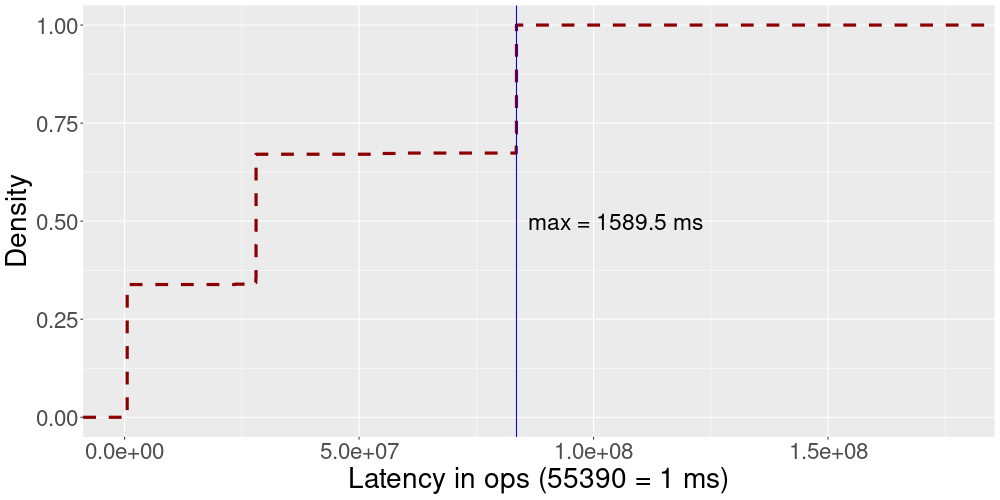}
  \caption{In-enclave Period Distribution} 
  
  \label{fig:enclave_length}
\end{figure}

We used this methodology to calculate the maximum in-enclave period before preemption, which is part of the calibration protocol. In Figure \ref{fig:enclave_length}, we show the cumulative distribution and highlight the maximum period achieved. The cumulative distribution curve shows enclave preemption is highly tri-modal, with $>99\%$ of exits happening in three small ranges. The blue line indicates the maximum enclave period, recorded at $\approx 1.58 s$. It also shows that nearly 30\% of in-enclave periods fall into the last range.

\begin{figure}
  \centering
  \includegraphics[width=\linewidth, height=5cm]{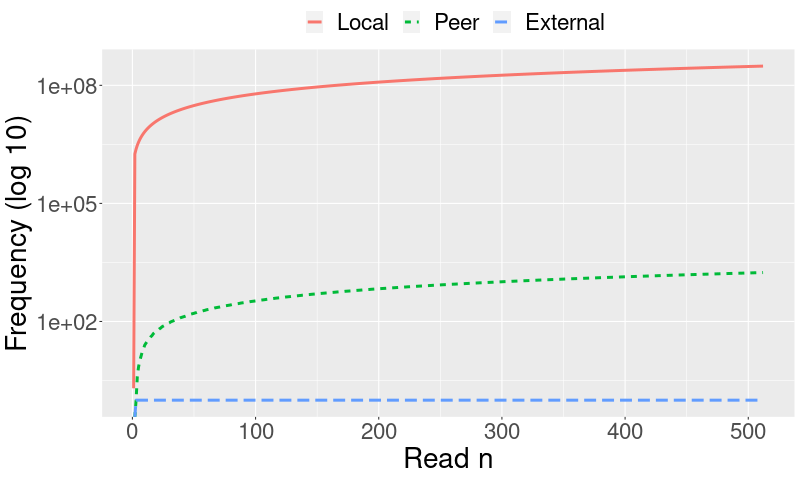}
  \caption{Frequency of time resources access -- Local vs Peer vs Remote} 
  
  \label{fig:aex_comb}
\end{figure}

\begin{figure}[htb]
  \centering
  \includegraphics[width=\linewidth, height=5.3cm]{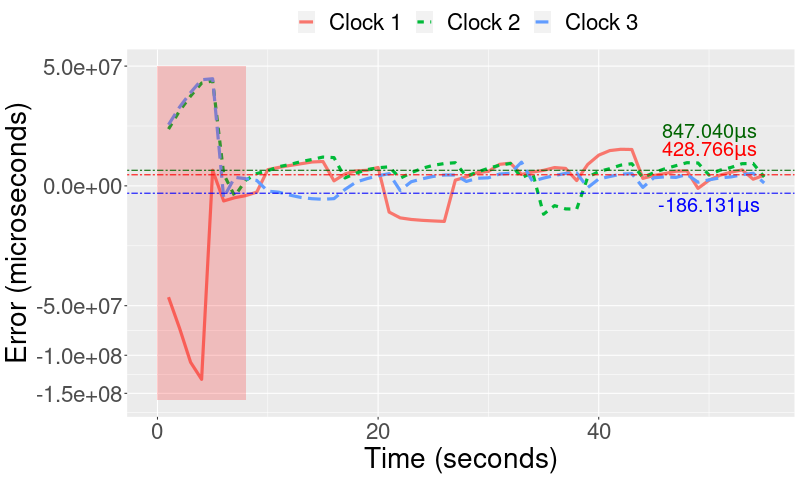}
  \caption{Inner clock error for nodes of a sample \textit{Triad} deployment with the calibration round.} 
  
  \label{fig:error}
\end{figure}

\subsection{Enclave Exit Rates under Optimal Kernel Configuration}
\label{sec:enclave_exits}

As referred to in Subsection \ref{sec:high_exits}, high transaction success rates are essential to maintain availability. In \textit{Triad}, an enclave exit delivered during a transaction will abort it. We have put in place kernel optimizations, as described in Subsubsection \ref{sec:tunning}, to allow for longer uninterrupted enclave execution.

\begin{figure}
  \centering
  \includegraphics[width=\linewidth, height=5cm]{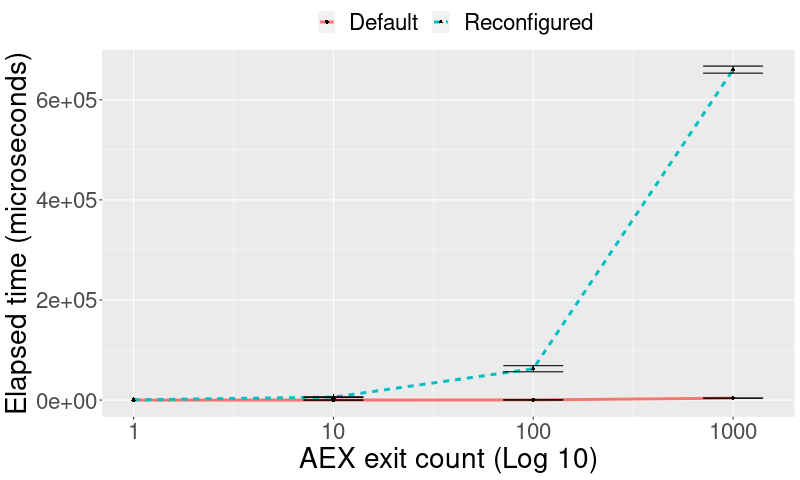}

  \caption{Elapsed time to reach a certain number of enclave exits -- Configured Kernel vs Default} 
    \label{fig:aex}
\end{figure}

In Figure~\ref{fig:aex}, we aim to isolate the kernel configuration factor by evaluating the respective configurations through a microbenchmark. The microbenchmark measures elapsed time for enclave exits in powers of 10. This experiment demonstrates the reduction of exits per second achieved with optimal kernel configuration.

With the default kernel configuration, the mean latency to reach 1000 exits is $\approx 3.93\times10^{3}$$ms$, while with the custom kernel configuration, we reach $\approx 6.24\times10^{5}$$ms$ -- two levels of magnitude difference.

\subsection{Round-Trip Time}

\begin{figure}[htb]
  \centering
  \includegraphics[width=\linewidth]{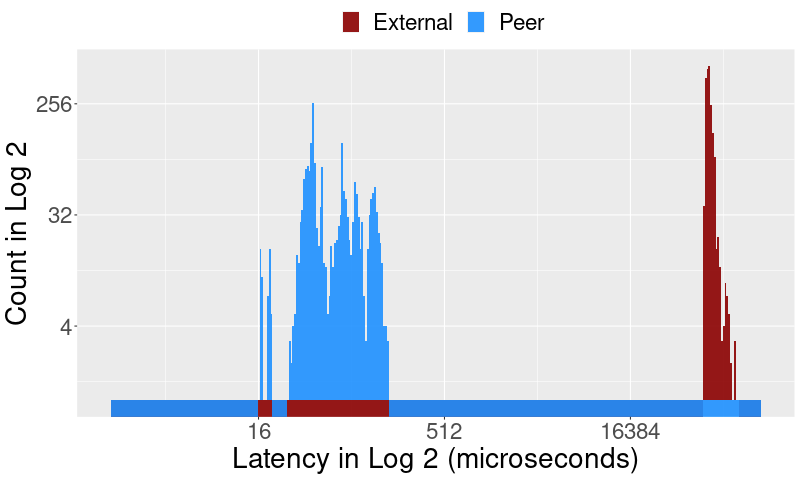}
  \caption{RTT for Peer and External Time Service timestamp reads.} 
  
  \label{fig:clock_reading_error}
\end{figure}

Here, we evaluate RTTs for peer and external time service timestamp reads. We made the peer measurements under a $1\ GB/s$ switched network. 

Figure \ref{fig:clock_reading_error} (note both axes are in $Log 2$ scale) depicts the distribution of measured clock reading error in $\mu$$s$ for 2048 samplings. We measured the mean clock reading error to be $69.493\ \mu $s, with a maximum measurement of $169.664\ \mu$s.
The external time service was a public Roughtime server. We measure its mean RTT at $\approx 142.766 $ ms.

In the experiment’s conditions, we also have found that the RTT for external time service presents a relatively low Standard Deviation, at $8486.725$. 
However, these conditions depend on a public IPv4 network, which may vary wildly. 


\subsection{Resource Access}

In this experiment, we evaluate with which rate each time a resource is accessed. We sample every $6\times10^5$ polling cycles from a $500\ s$ execution.

The timestamp count with peer server node origin has summed up to $\approx 1.73 \times 10^3$ occurrences, 
the timestamp count originated by local TSC reads summed up $\approx 3.07 \times 10^8$, as portrayed in Figure \ref{fig:aex_comb}. This result indicates that the vast majority of reads will have minimal impact on error growth, and the reconfiguration of kernel results in steady enclave-originated reads. We recorded one external read throughout the experiment. 

\subsection{Error Evaluation}

Figure \ref{fig:error} depicts a sample execution’s error between NTP updated Operating System-mediated clocks and \textit{Triad} cached timestamps. We represent the error in microseconds over the y-axis. Each of the \textit{Triad} server nodes has had their cached timestamps compared with that machine’s local Operating System clock to minimize error from the network.

Note that the graph area highlighted in red expresses the error during the calibration period. TSC’s drift generates a high accumulated error during that phase. The rate is, however, exceptionally constant. Calibration provides the system with the multiplication factor to correct the drift.

Most peer timestamps fall below the last cached timestamp due to latency and the brevity of interrupts in non-attack scenarios. Thus, they have the effect of simply elongating each slope. However, the samples for clocks 1 and 2 show pronounced spikes, which indicate longer interrupts taking place -- a typical sign of peer timestamp reads.

We measure the mean errors for each clock at $\approx 4.2\times10^2$$\mu$$s$, $\approx 8.4\times10^1$$\mu$$s$ and $\approx -1.8\times10^2$$\mu$$s$, all sitting within the sub-millisecond mark.

In our sample, we measure the instant maximum error modulus ($E(t_1) | \forall t \in time: |E(t_1)| > |E(t)|$) for each clock at $\approx 4.6\times10^3$  $\mu$$s$, $\approx 2.8\times10^4$  $\mu$$s$ and $\approx -1.9\times10^3$  $\mu$$s$. We, thus, recommend applications to maintain order at the tens of milliseconds magnitude. Events happening within that grain should be considered simultaneous, and the application must order them arbitrarily.

\section{Related Work}
\label{sec:related}

Vericount \cite{tople2018v} worked by using defunct native trusted time calls in between synchronous enclave exits and calculating in-enclave time by measuring the difference. It is vulnerable to provoking asynchronous exits, which would create indeterminate over-accounting of in-enclave time.

Aurora \cite{liang2018aurora} uses SMM to provide a collection of system services to TEE-based applications. A sufficiently dedicated adversary that controls all source hardware clocks while maintaining their monotonicity can attack that system’s time-keeping solution. Furthermore, the application is preempted on every read, significantly negatively impacting precision.

Works like Varys \cite{oleksenko2018varys}, and others employ several heuristics to detect or deter an ample class of SGX targeting side-channel attacks. These works use predictably timed operations to estimate time measurements like writes to main memory. They, however, do not maintain a long term notion of time.

S-FaaS \cite{alder2019s} uses sibling hyperthreads to count in-enclave periods, a more rustic approach, but similar to T-Lease \cite{trach2020t}, which provides the base time-keeping library we utilize as a building block. These works succeed in creating relative clocks capable of validating their measurements. They are, however, limited to the longest stay in an enclave. Furthermore, these works do not provide absolute timestamps.

In \cite{hamidy2023T3e}, the authors use TPM clocks with enclaves. While using TPMs is interesting as a seeding mechanism, the work does not present a strict solution for delay attacks. The work relies on the client’s eventual timing out from delayed reads, which would force new TPM timestamp reads since the usage of each timestamp is limited. Thus, if a client is attacked as not to time out of an operation, the delay in the delivery of the timestamp and, consequently, its error is arbitrarily high.

\section{Conclusion \& Future Work}
\label{sec:conclusion}

We have shown in this work a solution to the problem of trusted clocks in TEEs. We argue and demonstrate that, despite deliberate interruptions, multiple enclaves can form a notion of continuity of trusted timestamps. We show that our implementation, \textit{Triad}, abides by a set of rules that provides desirable properties to the client applications. These multiple enclaves must be setup in conjunction. As future work, for some architectures, such as AMD SEV-SNP \cite{sev2020strengthening} and Intel TDX \cite{cheng2023intel}, the service can provide a time reference for the whole virtual machine. We also aim at evaluating the second security layer with memory reads to support the CPU-based timing as well as the CPU-timing evalutation for different platforms.

\section{Acknowledgments} 
This work was supported by Cluster of Excellence “Centre for Tactile Internet with Human-in-the-Loop” (CeTI) of Technische Universität Dresden with Project ID 390696704, Federal Ministry of Education and Research of Germany in the programme of “Souverän. Digital. Vernetzt.” as project 6G-life with ID: 16KISK001K, and by European Commission through the Horizon Europe Research and Innovation program under Grant Agreement No. 101016577 (AI-SPRINT), 101092644 (NearData), and 101092646 (CloudSkin).

\end{document}